# Dark energy fluid with time-dependent, inhomogeneous equation of state

I. Brevik*,  O. G. Gorbunova** and A. V. Timoshkin**

*Department of Energy and Process Engineering, Norwegian University of Science and Technology, N-7491 Trondheim, Norway
**Tomsk State Pedagogical University, Tomsk, Russia

February 2007

Abstract

The four-dimensional Friedman flat universe, filled with an ideal fluid with a linear (oscillating) inhomogeneous equation of state (EoS) depending on time, is studied. The equations of motion are solved. It is shown that in some cases there appears a quasi-periodic universe, which repeats the cycles of phantom-type space acceleration. The appearance of future singularities resulting from various choices for the input parameters is discussed.

## 1. Introduction

As is known, the present universe is subject to an acceleration, which can be explained in terms of an ideal fluid (dark energy) weakly interacting with usual matter and which has an uncommon equation of state. The pressure of such an ideal dark energy fluid is negative. In the present work we study a model where there is an ideal fluid with an inhomogeneous equation of state, $p = w(t)\rho + \Lambda(t)$, in which the parameters $w(t)$ and $\Lambda(t)$ depend linearly on time. In another version, these parameters are oscillating in time. Ideal fluids with an inhomogeneous equation of state were introduced in [1] (see also examples discussed in [2]). We show that, depending on the choices for the input parameters, it is possible for the universe to pass from the phantom era to the non-phantom era, implying the appearance of a cosmological singularity. Also possible are cases of quasi-periodic changes of the energy density and of the Hubble constant, with the appearance of quasi-periodic singularities, and the appearance of cosmological singularities.

The particular kind of equation of state in the present paper is one alternative amongst a variety of possibilities, proposed recently to cope with the general dark energy problem (for an extensive review, see [3]). Different examples include imperfect equation of state [4], general equation of state [5], inhomogeneous equation of state [1], [6] including time-dependent viscosity as a special case [7], and multiple-Lambda cosmology [8].

In the next section, a linear inhomogeneous EoS ideal fluid in a FRW universe is studied. In section 3, an oscillating inhomogeneous EoS ideal fluid is investigated.

## 2. Inhomogeneous equation of state for the universe and its solution

Let us assume that the universe is filled with an ideal fluid (dark energy) obeying an inhomogeneous equation of state (see also Ref. [9]),

$$p = w(t)\rho + \Lambda(t), \tag{1}$$

where $w(t)$ and $\Lambda(t)$ depend on the time $t$. This equation of state, when $\Lambda(t) = 0$ but with $w(t)$ a function of time, was examined in Refs. [9] and [10].

Let us write down the law of energy conservation:

$$\dot\rho + 3H(\rho + p) = 0, \tag{2}$$

and Friedman's equation::

$$\frac{3}{\chi^2} H^2 = \rho, \tag{3}$$

where $\rho$ is the energy density, $p$ - the pressure, $H = \dfrac{\dot a}{a}$ - the Hubble parameter, $a(t)$ - the scale factor of the three-dimensional flat Friedman universe, and $\chi$ - the gravitational constant.

Taking into account equations (1), (2) and (3), we obtain the gravitational equation of motion :

$$\dot\rho + \sqrt{3}\chi[1 + w(t)]\rho^{3/2} + \sqrt{3}\chi\rho^{1/2}\Lambda(t) = 0. \tag{4}$$

Let us suppose in the following that both functions $w(t)$ and $\Lambda(t)$ depend linearly on time:

$$w(t) = a_1 t + b, \tag{5}$$

$$\Lambda(t) = ct + d. \tag{6}$$

This kind of behaviour may be the consequence of a modification of gravity (see [11] for a review).

Let $\Lambda(t) = 0$, $w(t) = a_1 t + b$. In this case the energy density takes the form:

$$\rho(t) = \frac{4(2a_1 + 1)^2}{3\chi^2} \cdot \frac{(a_1 t + b + 1)^{2/a_1}}{\left[(a_1 t + b + 1)^{\frac{1}{a_1} + 2} + S\right]^2}. \tag{7}$$

Hubble's parameter becomes:

$$H(t) = \frac{2}{3}(2a_1 + 1) \cdot \frac{(a_1 t + b + 1)^{\frac{1}{a_1}}}{(a_1 t + b + 1)^{\frac{1}{a_1} + 2} + S}, \tag{8}$$

where $S$ is an integration constant.

The time derivative of $H(t)$ becomes

$$\dot{H}(t) = \frac{2}{3}(2a_1+1)\cdot\left(1+\frac{1}{a_1}\right)\cdot(2a_1+b+1)^{\frac{2}{a_1}+1}\cdot\frac{\dfrac{a_1 S}{(a_1 t+b+1)^{\frac{1}{a_1}+2}}-1}{\left[(a_1 t+b+1)^{\frac{1}{a_1}+2}+S\right]^2}. \qquad (9)$$

The scale factor takes the following form:

$$a(t) = e^{\frac{2}{3}(2a_1+1)I},$$

where

$$I = \frac{(-1)^{a_1}}{(2a_1+1)S^{a_1}}\ln\left|(a_1 t+b+1)^{\frac{1}{a_1}}+S\right| - \frac{1}{(2a_1+1)S^{a_1}}\cdot$$

$$\cdot\sum_{k=0}^{a_1-1}\cos\frac{(a_1+1)(2k+1)}{2a_1+1}\cdot\pi\cdot\ln\left(\left[(a_1 t+b+1)^{\frac{2}{a_1}}+S\right]^2 - 2S(a_1 t+b+1)^{\frac{1}{a_1}}\cdot\cos\frac{2k+1}{2a_1+1}\pi+S^2\right)$$

$$+\frac{2}{(2a_1+1)S^a}\cdot \qquad (10)$$

$$\cdot\sum_{k=0}^{a_1-1}\sin\frac{(a_1+1)(2k+1)}{2a_1+1}\cdot\pi\cdot arctg\,\frac{(a_1 t+b+1)^{\frac{1}{a_1}}-S\cdot\cos\dfrac{2k+1}{2a_1+1}\cdot\pi}{S\cdot\sin\dfrac{2k+1}{2a_1+1}\cdot\pi},$$

$1 \leq a_1 \leq 2a_1 - 1$.

At $t_1 = -\dfrac{b+1}{a_1}$ or $t_2 = \dfrac{1}{a_1}\left(\dfrac{Sa_1}{a_1+1}\right)^{\frac{1}{2+\frac{1}{a_1}}} - \dfrac{b+1}{a_1}$, one has $\dot{H}=0$. With $a_1 > 0$, $b > -1$ and $t < t_2$ one gets $\dot{H} > 0$, that is, the accelerating universe is in the phantom phase (see, for example, [12]), and with $t > t_2$, one gets $\dot{H} < 0$, the universe is in the non-phantom phase. At the moment when the universe passes from the phantom to the non-phantom era, Hubble's parameter equals

$$H_m = \frac{2}{3S}\sqrt[2a_1+1]{a_1(a_1+1)^{2a_1}}. \qquad (11)$$

In the phantom phase $\dot{\rho} > 0$ the energy density grows; in the non-phantom phase $\dot{\rho} < 0$ the energy density decreases. However, the derivative of the scale factor $\dot{a} > 0$, therefore the universe expands. Note, as has been shown in Ref. [9], that in the phantom phase the entropy may become negative. If $t \to +\infty$, then $H(t)$ and $\rho(t) \to 0$, so that the phantom energy decreases.

A graph of the function $H$ versus time $t$, in the case when $a_1$ is an odd number, is shown in Fig. 1.

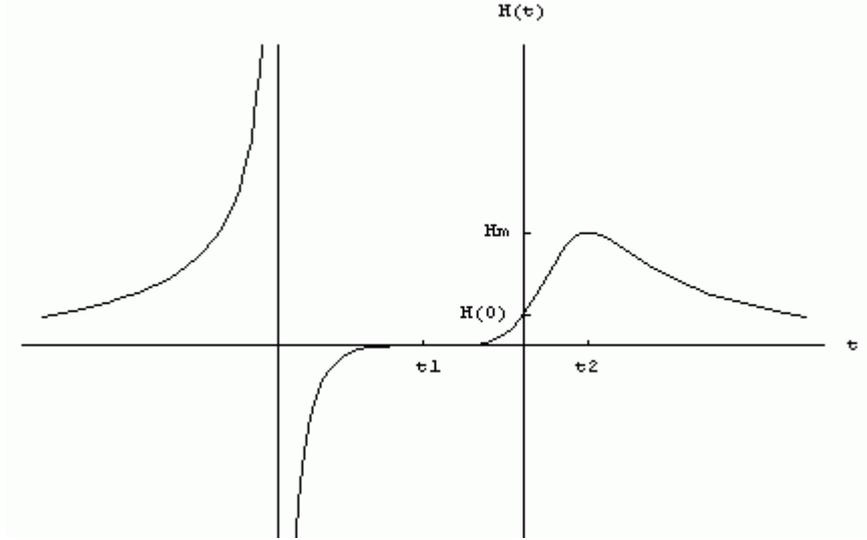

$$H(0) = \frac{\frac{2}{3}(2a_1+1)}{(b+1)^2 + S}$$

*Fig. 1*

The simultaneous divergence of $\rho(t)$ and $H(t)$ (cosmological singularity) appears at

$$t = \frac{1}{a_1} S^{\frac{a_1}{2a_1+1}} - \frac{b+1}{a_1}$$ (for a classification of singularities, see [13]).

For even values of $a_1$, a graphical representation of $H(t)$ is given in Fig. 2:

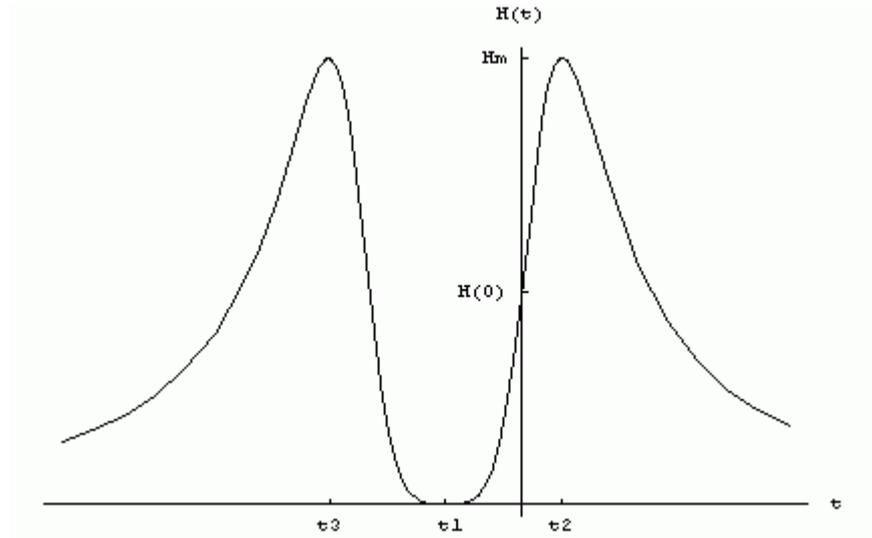

$$t_3 = -\frac{1}{a_1}\left(\frac{a_1 S}{a_1+1}\right)^{2+\frac{1}{a_1}} - \frac{b+1}{a_1}.$$

*Fig. 2*

Here is repeated the process of the universe passing from a phantom phase to a non-phantom phase.

Let us examine the case when $w(t) = 0$, $\Lambda(t) = ct + d$. For the energy density we then obtain

$$\rho(\xi) = \frac{4}{3\chi^2 \xi^2 \left(1 + \dfrac{\frac{1}{3}\gamma}{\xi^3 - \frac{2}{3}\sqrt{\gamma}\xi^{\frac{3}{2}}}\right)^2},\qquad(12)$$

where $\gamma = \dfrac{4}{3c\chi}$, $\xi = \dfrac{3}{4}\chi^2(ct + d)$.

Hubble's parameter takes the form:

$$H(\xi) = \cfrac{2}{3\xi \cdot \left(1 + \cfrac{\frac{1}{3}\gamma}{\xi^3 - \frac{2}{3}\sqrt{\gamma}\xi^{\frac{3}{2}}}\right)}, \qquad (13)$$

and its derivative becomes:

$$\dot{H}(\xi) = -\frac{2}{3} \cdot \frac{\xi^4 - \frac{4}{3}\sqrt{\gamma}\xi^{\frac{5}{2}} - \frac{2}{9}\gamma\xi + \frac{1}{9}\gamma^{\frac{3}{2}} \cdot \xi^{-\frac{1}{2}}}{\left(\xi^3 - \frac{2}{3}\gamma\xi^{\frac{3}{2}} + \frac{1}{3}\gamma\right)^2}. \qquad (14)$$

The scale factor is given by the expression:

$$a(\xi) = \frac{S \cdot \sqrt[9]{\left(\xi^3 - \frac{2}{3}\sqrt{\gamma}\xi^{\frac{3}{2}} + \frac{1}{3}\gamma\right)^2}}{e^{\frac{\sqrt{2}}{3} \cdot arctg \frac{3\xi^{3/2} - \sqrt{\gamma}}{2\sqrt{\gamma}}}}, \qquad (15)$$

where S is an integration constant.

The derivative $\dot{H} > 0$ if $\xi_1 < \xi < \xi_2$, and we obtain the phantom phase of the universe. When $\xi < \xi_1$ and $\xi > \xi_2$ the derivative $\dot{H} < 0$ and we obtain the non-phantom phase.

If $\xi \to 0$ $(t \to -\frac{d}{c})$, the functions $H(\xi)$ and $\rho(\xi)$ simultaneously approach zero. When $\xi < \xi_2$, $\dot{\rho} > 0$ the energy density grows, and there occurs an expansion of the universe. In the case when $\xi > \xi_2$, $\dot{\rho} < 0$, $\dot{a} < 0$, there occurs a compression of the universe.

A graphical representation of $H(\xi)$ is given in Fig. 3:

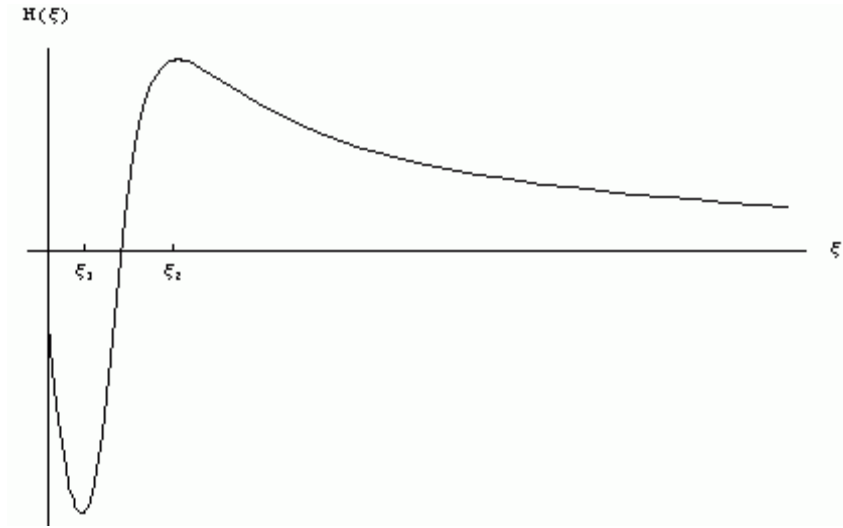

$\xi_1 = 0{,}757 \cdot \gamma^{\frac{1}{3}}, \xi_2 = 1{,}272 \cdot \gamma^{\frac{1}{3}}$

*Fig.3*

Now let us examine $w(t) = b$, $\Lambda(t) = ct + d$. In this case we have:

$$\rho(\xi) = \frac{4}{3\chi^2 \xi^2 (1+b)^2 \left(1 + \frac{\frac{1}{3}\gamma}{\xi^3 - \frac{2}{3}\sqrt{\gamma}\xi^{\frac{3}{2}}}\right)^2}, \qquad (16)$$

$$H(\xi) = \frac{2}{3\xi(1+b)\cdot\left(1 + \frac{\frac{1}{3}\gamma}{\xi^3 - \frac{2}{3}\sqrt{\gamma}\xi^{\frac{3}{2}}}\right)}, \qquad (17)$$

$b \neq -1$.

If $b > -1$, the energy density and the Hubble parameter change analogously to the case $w(t) = 0$. But if $b < -1$, then with $\xi < \xi_2$ the energy density diminishes $\dot{\rho} < 0$, and $\dot{a} < 0$, and there occurs a compression of the universe. When $\dot{\rho} > 0$, $\xi > \xi_2$, there occurs an expansion of the universe.

Assume now that $w(t) = at + b$, and $\Lambda(t) = d$. Then the energy density takes the form:

$$\rho(t) = \frac{d}{3(a_1 t + b + 1)} \cdot \left( \frac{Z_{-\frac{1}{3}}\left[\frac{\chi}{a_1}\cdot\sqrt{\frac{d}{3}}(a_1 t + b + 1)^{\frac{3}{2}}\right]}{Z_{\frac{2}{3}}\left[\frac{\chi}{a_1}\sqrt{\frac{d}{3}}(a_1 t + b + 1)^{\frac{3}{2}}\right]} \right)^2, \qquad (18)$$

and the Hubble parameter becomes:

$$H(t) = \frac{\chi}{3}\frac{\sqrt{d}}{\sqrt{a_1 t + b + 1}} \cdot \frac{Z_{-\frac{1}{3}}\left[\frac{\chi}{a_1}\sqrt{\frac{d}{3}}(a_1 t + b + 1)^{\frac{3}{2}}\right]}{Z_{\frac{2}{3}}\left[\frac{\chi}{a_1}\sqrt{\frac{d}{3}}(a_1 t + b + 1)^{\frac{3}{2}}\right]}, \qquad (19)$$

where $Z_\nu = C_1 J_\nu + C_2 Y_\nu$, i.e. the general solution of Bessel's equation. Moreover, $C_1, C_2$ are arbitrary constants, and $t > -\frac{b+1}{a_1}$.

The derivative equals:

$$\dot{H}(t) = -\frac{d\chi^2}{6}\left[1 + \left(\frac{Z_{-\frac{1}{3}}}{Z_{\frac{2}{3}}}\right)^2\right]. \qquad (20)$$

Since $d > 0$, we have $\dot{H} < 0$, and we get the non-phantom phase of the universe. The functions $\rho(t)$ and $H(t)$ change quasi-periodically (Fig. 4). Cosmological singularities appear also quasi-periodically. The energy density and the Hubble parameter simultaneously approach infinity. That is, we have a singularity of the type Big Rip.

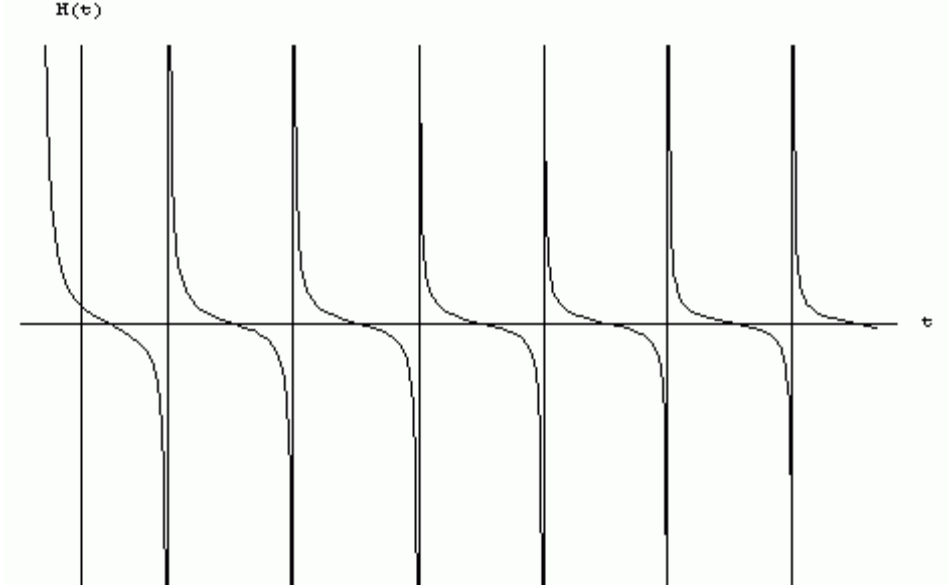

*Fig.4*

$$H(0) = \frac{\chi}{3}\sqrt{\frac{d}{b+1}} \cdot \frac{Z_{-1/3}\left[\frac{\chi}{a_1}\sqrt{\frac{d}{3}}(b+1)^{3/2}\right]}{Z_{2/3}\left[\frac{\chi}{a_1}\sqrt{\frac{d}{3}}(b+1)^{3/2}\right]}. \tag{21}$$

### 3. Inhomogeneous oscillating equation of state

Instead of assuming the form (5) and (6) for the time-dependent parameters, one might assume that there is an oscillating dependence on time. Let us investigate the following form: $w(t) = -1 + \omega_0 \cos \omega t$. From equations (3) and (4) we get

$$H = \frac{2\omega}{3(\omega_1 + \omega_0 \sin \omega t)}, \tag{22}$$

where $\omega_1$ is an integration constant. If $|\omega_1| < \omega_0$, the denominator can in this case be zero, what implies a future cosmological singularity. But if $|\omega_1| > \omega_0$, the singularity is absent.

In view of the fact that (cf. [10])

$$\dot{H} = -\frac{2\omega^2 \omega_0 \cos \omega t}{3(\omega_1 + \omega_0 \sin \omega t)^2}, \tag{23}$$

we see that with $\omega_0 \cos \omega t < 0$ ($\omega_0 \cos \omega t > 0$) the universe is located in the phantom (non-phantom) phase, corresponding respectively to $\dot{H} > 0$ ($\dot{H} < 0$). If the oscillation period of the universe is large, it is possible to have a unification of inflation and phantom dark energy [12]. The density of dark energy takes the form

$$\rho(t) = \frac{4\omega^2}{3\chi^2(\omega_1 + \omega_0 \sin \omega t)^2}, \qquad (24)$$

this being a periodic function so that the universe oscillates between the phantom and non-phantom eras.

Let us assume now that $\Lambda(t) \neq 0$. For simplicity we take $\Lambda(t) = \Lambda_0 \cos \omega t$, i.e. a periodic function. If $\Lambda_0 < 0$, equation (4) has the following solution:

$$\frac{\sqrt{\rho(t)} + \sqrt{\frac{|\Lambda_0|}{\omega_0}}}{\sqrt{\rho(t)} - \sqrt{\frac{|\Lambda_0|}{\omega_0}}} = \exp\left[\sqrt{3\chi^2|\Lambda_0|\omega_0}\left(\frac{\sin \omega t}{\omega} + C_1\right)\right], \qquad (25)$$

where $C_1$ is an integration constant.

Finally, we obtain for the energy density:

$$\rho(t) = \left\{ \frac{\sqrt{\frac{|\Lambda_0|}{\omega_0}}\left\{\exp\left[\sqrt{3\chi^2|\Lambda_0|\omega_0}\left(\frac{\sin \omega t}{\omega} + C_1\right)\right]\right\} + \sqrt{\frac{|\Lambda_0|}{\omega_0}}}{\exp\left[\sqrt{3\chi^2|\Lambda_0|\omega_0}\left(\frac{\sin \omega t}{\omega} + C_1\right)\right] - 1} \right\}^2. \qquad (26)$$

The Hubble parameter becomes, according to (3),

$$H(t) = \sqrt{\frac{\chi^2 \rho(t)}{3}}. \qquad (27)$$

At the moments when the denominator of (26) is zero, the energy density diverges. This corresponds to a future cosmological singularity. Depending on the choice of parameters in the equation of state for the dark energy, $H(t)$ can thus correspond to either a phantom, or a non-phantom, universe. In both cases the universe expands with (quintessential or super) acceleration.

## 4. Summary

In this work we have studied a model of the universe in which there is a linear inhomogeneous equation of state, with a linear or oscillating dependence on time. The consequences of various choices of parameters in the linear functions are examined: there may occur a passage from the non-phantom era of the universe to the phantom era, resulting in an expansion and a possible appearance of singularities. In the absence of inhomogeneous terms it is possible to have a repetition of the passage process. When the universe goes from the phantom era to the non-phantom era with expansion one may avoid singularities, or there may appear singularities, but the passage occurs without repetition.

The presence of a linear inhomogeneous term in the equation of state leads either to a compression of the universe in the evolution process or to a quasi-periodic change in the energy density and in the Hubble parameter, and also to a quasi-periodic appearance of singularities. By this the universe either passes into the non-phantom era, or stays within the same era as it was originally. Thus, the universe filled with an inhomogeneous time-dependent equation-of-state ideal fluid may currently be in the acceleration epoch of quintessence or phantom type. Moreover, it is easy to see that the effective value of the equation-of-state parameter may easily be adjusted so as to be approximately equal to -1 at present, what corresponds to current observational bounds.

More precise determination of the cosmological parameters and, in particular, the derivative of $w(t)$, can answer the question of how an ideal fluid of the considered type is realistic.

**Acknowledgment**

We thank Sergei Odintsov for very valuable information.

This work is supported by Grant the Scientific School LRSS № 4489.2006.02., and also by Grant RFBR 06-01-00609.